\documentclass[aps,12pt,showpacs]{revtex4}
\usepackage{epsfig}

\begin{document}

\newcommand{\half}{\frac12}
\newcommand{\vare}{\varepsilon}
\newcommand{\eps}{\epsilon}
\newcommand{\pr}{^{\prime}}
\newcommand{\ppr}{^{\prime\prime}}
\newcommand{\pp}{{p^{\prime}}}
\newcommand{\hp}{\hat{\bfp}}
\newcommand{\hpp}{\hat{\bfpp}}
\newcommand{\hq}{\hat{\bfq}}
\newcommand{\rqq}{{\rm q}}
\newcommand{\rx}{{\rm x}}
\newcommand{\rp}{{\rm p}}
\newcommand{\rpp}{{{\rm p}^{\prime}}}
\newcommand{\rk}{{\rm k}}
\newcommand{\bfp}{{\bf p}}
\newcommand{\bfpp}{{\bf p}^{\prime}}
\newcommand{\bfq}{{\bf q}}
\newcommand{\bfx}{{\bf x}}
\newcommand{\bfk}{{\bf k}}
\newcommand{\bfz}{{\bf z}}
\newcommand{\bphi}{{\mbox{\boldmath$\phi$}}}
\newcommand{\balpha}{{\mbox{\boldmath$\alpha$}}}
\newcommand{\bsigma}{{\mbox{\boldmath$\sigma$}}}
\newcommand{\bomega}{{\mbox{\boldmath$\omega$}}}
\newcommand{\bvare}{{\mbox{\boldmath$\varepsilon$}}}
\newcommand{\intzo}{\int_0^1}
\newcommand{\intinf}{\int^{\infty}_{-\infty}}
\newcommand{\ka}{\kappa_a}
\newcommand{\kb}{\kappa_b}
\newcommand{\lbr}{\langle}
\newcommand{\rbr}{\rangle}
\newcommand{\ThreeJ}[6]{
        \left(
        \begin{array}{ccc}
        #1  & #2  & #3 \\
        #4  & #5  & #6 \\
        \end{array}
        \right)
        }
\newcommand{\SixJ}[6]{
        \left\{
        \begin{array}{ccc}
        #1  & #2  & #3 \\
        #4  & #5  & #6 \\
        \end{array}
        \right\}
        }
\newcommand{\NineJ}[9]{
        \left\{
        \begin{array}{ccc}
        #1  & #2  & #3 \\
        #4  & #5  & #6 \\
        #7  & #8  & #9 \\
        \end{array}
        \right\}
        }
\newcommand{\Dmatrix}[4]{
        \left(
        \begin{array}{cc}
        #1  & #2   \\
        #3  & #4   \\
        \end{array}
        \right)
        }
\newcommand{\cross}[1]{#1\!\!\!/}
\newcommand{\beq}{\begin{equation}}
\newcommand{\eeq}{\end{equation}}
\newcommand{\beqn}{\begin{eqnarray}}
\newcommand{\eeqn}{\end{eqnarray}}

%
%
\title{
Two-loop self-energy correction in high-$\mathbf{Z}$ hydrogen-like ions }
\author{
V. A. Yerokhin$^{1,2}$, P. Indelicato$^2$, and V. M. Shabaev$^{1}$ }
\affiliation{ $^1$ Department of Physics, St.~Petersburg State University,
Oulianovskaya 1, St.~Petersburg 198504, Russia\\
$^2$ Laboratoire Kastler-Brossel, Unit\'e mixte de l'\'Ecole Normale
Sup\'erieure, du CNRS et l'Universit\'e P. et M. Curie, Case 74, 4 place
Jussieu, F-75252, Cedex 05, France\\ }

\begin{abstract}
A complete evaluation of the two-loop self-energy diagrams to all orders in
$Z\alpha$ is presented for the ground state of H-like ions with $Z\ge 40$.

\end{abstract}

\date{ Feb. 28, 2003}
\maketitle

%

The calculation of the two-loop self-energy correction to the Lamb shift is
at present one of the most challenging problems in bound-state QED. Until
very recently, this project has been addressed to mainly within the
$Z\alpha$-expansion approach. In it, the two-loop self-energy contribution is
represented as an expansion over $Z\alpha$ and $\ln(Z\alpha)$,
\beqn  \label{aZexp}
F(Z\alpha) &=& B_{40}+ (Z\alpha)B_{50} + (Z\alpha)^2 \Bigl[
  L^3B_{63}
\nonumber \\ &&
  +L^2B_{62} +
  L\,B_{61} + B_{60} \Bigr]+ \cdots \,,
\eeqn
where $F(Z\alpha) = \Delta E /[m (\alpha/\pi)^2 (Z\alpha)^4]$, $L =
\ln(Z\alpha)^{-2}$. Whereas the lowest-order term $B_{40}$ has been known for
a long time, calculations of higher-order contributions have not been
accomplished until recently. The correction $B_{50}$ was found to be
surprisingly large \cite{Pachucki94,Eides95}, $B_{50} = -24.27$. This result
has significantly changed the theoretical prediction for the Lamb shift in
hydrogen and resolved the disagreement with the experimental value existing
at that time. The leading logarithmic term $B_{63}$ was derived first in
\cite{Karshenboim93} and later confirmed in \cite{Manohar00,Pachucki01}. (It
should be noted that the first evaluation \cite{Karshenboim93}, while
yielding the right result, is not completely correct, as will be discussed
below.) The two remaining logarithmic corrections $B_{62}$ and $B_{61}$ have
also been elaborated lately by Pachucki \cite{Pachucki01}. Again, as in order
$(Z\alpha)^5$, the result obtained turned out to be surprisingly large. The
numerical value of $B_{61}$ is $50.3$, which reverses the sign of the overall
logarithmic contribution for hydrogen. This indicates that the convergence of
the $Z\alpha$ expansion for the two-loop self-energy correction is remarkably
slow, and a conclusion has been drawn in \cite{Pachucki01} that a numerical
evaluation with Dirac-Coulomb propagators is desirable even for hydrogen.

\begin{figure}
\centerline{ \mbox{ \epsfxsize=0.47\textwidth \epsffile{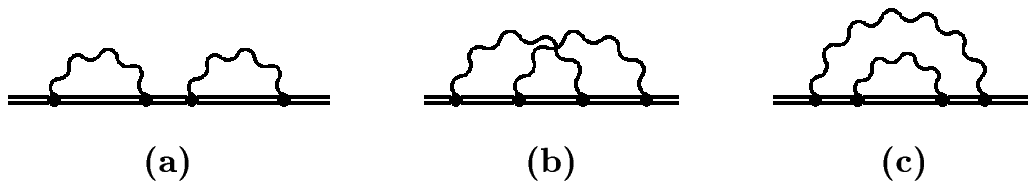} }}
\vskip0.5cm \caption{Two-loop self-energy diagrams. Double line indicates an
electron propagating in the Coulomb nuclear field. It is understood that the
corresponding mass counterterms are subtracted from the diagrams.
\label{sese}}
\end{figure}

The calculation of the two-loop self-energy diagrams (Fig.~\ref{sese})
without an expansion in $Z\alpha$ started with the irreducible contribution
of the diagram (a) (known also as the loop-after-loop (LAL) correction),
which is by far the simplest part of the total set. Such an evaluation was
first accomplished in \cite{Mitrushenkov95} for high-$Z$ ions, and later in
\cite{Mallampalli98} for all ions, including hydrogen. The latter
investigation demonstrated a rather peculiar behaviour of the LAL correction
in the low-$Z$ region. It was shown that for hydrogen its actual value was of
a different sign and magnitude than the value based on first two terms of the
$Z\alpha$ expansion. In addition, a different result was found in
\cite{Mallampalli98} for the leading logarithmic contribution $B_{63}$ as
compared to the analytical evaluation \cite{Karshenboim93}. (We note that in
the latter work the $B_{63}$ term was evaluated for the whole set of two-loop
self-energy diagrams. However, it was argued that it originated from the LAL
contribution only.) As a result, a question was raised in
\cite{Mallampalli98} about the possibility that the $Z\alpha$ expansion for
the two-loop self-energy could be inadequate even for hydrogen. This
speculation attracted attention and several investigations followed. The
subsequent numerical calculation \cite{Goidenko99} claimed to be compatible
with the analytical result. However, the third numerical evaluation by one of
us \cite{Yerokhin00lal} confirmed the first result \cite{Mallampalli98}. At
the same time, the total value of the $B_{63}$ contribution was confirmed
independently by several groups, e.g., in \cite{Manohar00}. To throw light on
this intricate situation we performed \cite{Yerokhin01lal} an analytic
calculation of the $B_{63}$ term separately for the LAL correction and found
agreement both with our previous numerical result and with that of
\cite{Mallampalli98}. Our conclusion was that the LAL correction provided an
additional cubed logarithmic contribution that had been omitted in the
original analytical calculation \cite{Karshenboim93}. However, this
additional term vanishes when the whole set of two-loop self-energy diagrams
is taken into account. Recently, analogous additional terms were reported for
the leading logarithmic contribution for $P$ states \cite{Jentschura02}.

The evaluation of the remaining contributions in Fig.~\ref{sese} is by far
more difficult. These contributions are: the reducible part of the diagram
(a), the overlapping diagram (b), and the nested diagram (c). The first
attempt to evaluate them to all orders in $Z\alpha$ was made by Mallampalli
and Sapirstein \cite{Mallampalli98a}. In that work, the contribution of
interest was rearranged in 3 parts, referred to by the authors as the "$M$",
"$P$", and "$F$" terms. (We will discuss this separation in more detail
below.) Mallampalli and Sapirstein calculated only the $M$ and $F$ terms,
while the $P$ term was left out since a new numerical technique had to be
developed for its computation. In addition, since the numerical procedure
turned out to be very time consuming, the actual calculation of the $M$ term
was carried out only for two ions, uranium and bismuth. Subsequently, in the
investigation by two of us \cite{Yerokhin01sese} we accomplished the
computation of the remaining $P$ term for $Z=83$, 90, and 92, which formally
completed the calculation of the two-loop self-energy. However, as we will
see, the rearrangement of the whole correction into the $M$, $P$, and $F$
terms is artificial since all the three are divergent. A proper treatment
should include these terms simultaneously. In addition, more than two points
(in $Z$) are needed in order to analyze the $Z$ dependence of the correction
and to compare it with the known terms of the $Z\alpha$ expansion. All these
issues are addressed to in the present investigation.

Let us now turn to the evaluation of the two-loop self-energy diagrams. The
first problem to be solved is the separation of ultraviolet (UV) divergences.
The standard method of renormalization in QED is developed for diagrams
involving only free-electron propagators, treating them in momentum space.
Therefore, our strategy is to subtract similar diagrams with electron
propagators containing zero or one interaction with the nuclear Coulomb field
in order to make the corresponding point-by-point difference UV finite. The
subtracted diagrams can be then evaluated in momentum space or in the mixed
momentum-coordinate representation. For the first-order self-energy, this
approach was first implemented in \cite{Snyderman91Blundell91}. The situation
is much more difficult in the case of the two-loop self-energy. Here, for the
first time, we encounter the overlapping UV divergences. For example, the
diagram in Fig.~\ref{sese}(b) can be considered as consisting of two
overlapping vertex subdiagrams, each of which is UV divergent. The presence
of the overlapping divergences makes the structure of subtraction terms much
more elaborate than that in the first order. Moreover, some of these terms
contain both bound-electron propagators and UV-divergent subdiagrams. Such
situation had never been encountered before, and a new numerical technique
had to be developed for the evaluation of these subtraction terms.

\begin{figure}
\centerline{ \mbox{ \epsfxsize=0.47\textwidth \epsffile{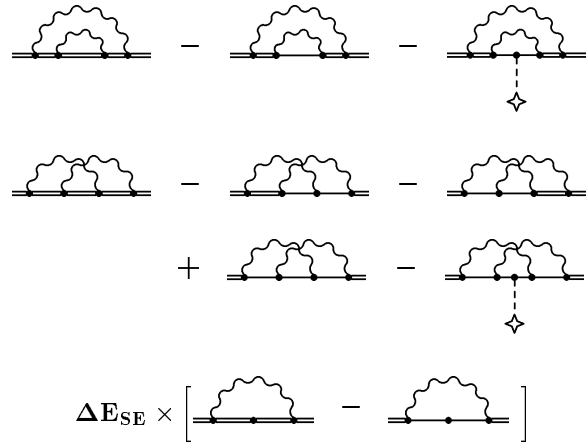} }}
\vskip0.5cm \caption{Diagrammatic representation of the $M$ term. The dashed
line denotes the interaction with the Coulomb nuclear field. $\Delta E_{\rm
SE }$ is the first-order self-energy correction. \label{mterm}}
\end{figure}

\begin{figure}
\centerline{ \mbox{ \epsfxsize=0.47\textwidth \epsffile{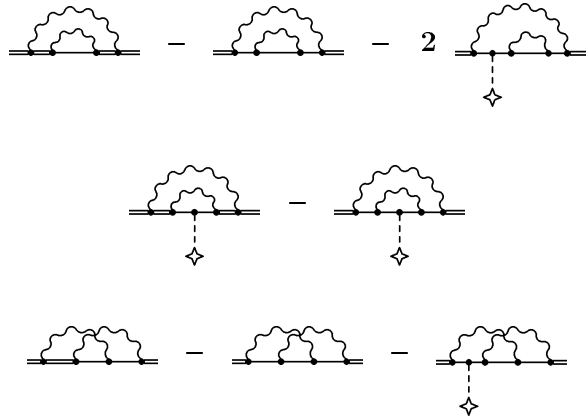} }}
\vskip0.5cm \caption{Diagrammatic representation of the $P$ term. The last
part should be counted twice, accounting for two equivalent terms.
\label{pterm}}
\end{figure}

\begin{figure}
\centerline{ \mbox{ \epsfxsize=0.47\textwidth \epsffile{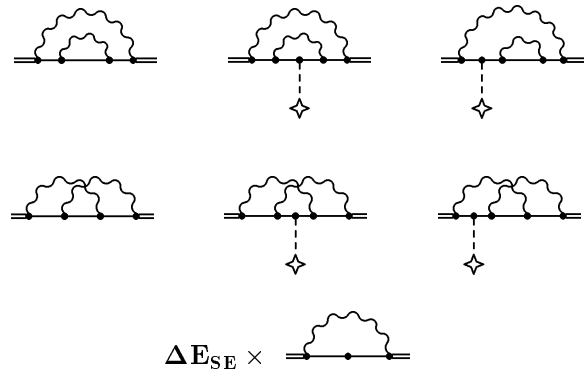} }}
\vskip0.5cm \caption{Diagrammatic representation of the $F$ term. The last
diagram on the right in the first two rows should be counted twice,
accounting for two equivalent diagrams.  \label{fterm}}
\end{figure}

Following \cite{Mallampalli98a}, we rearrange the contribution of the
diagrams in Fig.~\ref{sese} in 4 parts: the LAL part, the $M$,  $P$, and $F$
terms. The LAL correction is defined by the irreducible part of the diagram
(a). Since its evaluation is relatively easy and has been performed by
several groups, we do not discuss it here. The $M$ term is diagrammatically
represented by Fig.~\ref{mterm}. It consists of 3 parts originating from the
nested diagram, the overlapping diagram, and the reducible part of the
diagram (a). The subtractions in the $M$ term are chosen so that each of
these 3 parts is separately UV finite. Next, we should account for the
subtracted terms. Those that contain only free-electron propagators can be
treated in momentum space using the standard Feynman-parametrization
technique. For those that involve bound-electron propagators, we introduce
additional subtractions that remove the {\it overlapping} UV divergences.
This is graphically represented by Fig.~\ref{pterm}. The corresponding
contribution is referred to as the $P$ term. It consists of 3 parts, each
containing only single UV-divergent subgraphs. Taking the first part as an
example, we see that the difference shown in the figure is UV divergent only
due to the inner self-energy loop, while the divergence due to the outer
self-energy loop is canceled. Finally, we collect all terms we have
subtracted and denote them as the $F$ term depicted in Fig.~\ref{fterm}. It
consists of Feynman diagrams that contain free-electron propagators only.

We should also mention the infrared (IR) reference-state divergences that are
present in the $M$ and $P$ terms. These singularities can occur in
bound-state QED calculations when energies of the intermediate states in the
spectral decomposition of electron propagators coincide with the
valence-state energy. An analysis given, e.g., in \cite{Mallampalli98a} shows
that the IR-divergent terms cancel each other in the sum of the $M$ and the
$P$ term. To sum up our discussion of divergences, we separately write
divergent contributions to the terms under consideration:
\beqn
 \Delta E_{M} &=& \Delta E_{M}^{f}- \Delta E_{\rm IR}\,, \\
 \Delta E_{P} &=& \Delta E_{P}^{f}+ \Delta E_{\rm IR}
     + L^{(1)} \Delta E_{\rm SE}^{(2+)}\,, \\
 \Delta E_{F} &=& \Delta E_{F}^{f}
     + B^{(1)} \Delta E_{\rm SE}^{(2+)}\,,
\eeqn
where the index $f$ labels finite contributions, $\Delta E_{\rm IR}$ is the
IR-divergent contribution, $L^{(1)}$ and $B^{(1)}$ are the one-loop
renormalization constants that fulfill the Ward identity $L^{(1)} =
-B^{(1)}$, and $\Delta E_{\rm SE}^{(2+)}$ is the many-potential part of the
one-loop self-energy correction.

We now turn to the numerical evaluation of these terms. It was carried out in
the Feynman gauge. The $P$ term was evaluated along the lines described in
detail in our previous investigation \cite{Yerokhin01sese}. The calculation
of the $F$ and $M$ terms was performed in a way, in many respects similar to
that of \cite{Mallampalli98a}. The details of the calculation will be
published elsewhere. Here, we focus on major novel features of our evaluation
as compared to \cite{Mallampalli98a}. The first point is a different
treatment of the reference-state IR divergences. In \cite{Mallampalli98a},
they were regulated by altering the valence energy $\vare_a$ to
$\tilde{\vare}_a = \vare_a(1-\delta)$. The actual calculations were performed
keeping a finite regulator $\delta$, and the limit $\delta \to 0$ was
evaluated numerically. According to our experience, that approach, while
being technically easy to handle, does not allow to control the accuracy of
the computation effectively. In our approach, we introduce some subtractions
in order to make the terms under consideration finite, separating IR
divergences in the form $\Delta E_{\rm IR}$. The divergent contributions
cancel each other explicitly in the sum, and we can perform the whole
computation without introducing any actual IR regulator. However, in order to
allow the term-by-term comparison with the previous evaluation
\cite{Mallampalli98a}, we performed our calculations with the regulator
$\delta$ as well.

The second new feature of our approach is a different procedure employed for
the double summation over the partial waves in the $M$ term. In
\cite{Mallampalli98a}, the photon angular momenta $l_1$ and $l_2$ were chosen
as independent expansion parameters. We found it technically more convenient
to employ for this purpose the absolute values of the relativistic angular
parameter $\kappa$ of two electron propagators, $|\kappa|= j+1/2$. Thus, we
turn the nested and overlapping contributions to the $M$ term into tables of
values $X_{|\kappa_1|,|\kappa_2|}$. Next, we perform a resummation of the
table: $X_{|\kappa_1|,|\kappa_2|} \to Y_{ij}$, where $i = \left|
|\kappa_1|-|\kappa_2| \right|$, $j = |\kappa_1|+|\kappa_2|$. Finally we sum
up the table: first, we fix $i$ and extrapolate the sum over $j$ to infinity,
and then sum over $i$ and estimate the tail of the expansion.

Now we discuss the computer time necessary for the evaluation of the $M$
term. In the previous evaluation by Mallampalli and Sapirstein, a total time
of 7323 h was required for a given value of $Z$. In our numerical approach,
the typical time of the evaluation of one element $X_{\kappa_1\kappa_2}$ is
about 1 h for the IBM PWR3 processor with 350 MHz, both for the nested and
the overlapping diagram. The typical number of elements for a given $Z$ was
440 for the nested diagram and 320 for the overlapping diagram. This shows
that the time consumption in our numerical procedure is smaller than that of
\cite{Mallampalli98a}, although it is still very large.

In Table I we present the numerical results for finite parts of the $M$, $P$,
and $F$ terms. The table shows that our numerical values for the LAL and $F$
terms agree very well with the ones from \cite{Mallampalli98,Mallampalli98a}
but there is a significant deviation for the $M$ term. More specifically, our
calculation for $Z=92$ yields $-$2.137, 4.679(2), and $-$3.837(2) for the
reducible, nested, and overlapping contributions to the $M$ term,
respectively. These results should be compared correspondingly with $-$2.137,
4.669(5), and $-$4.387(5) from \cite{Mallampalli98a}. We see that the leading
source of discrepancy is the overlapping diagram. Taking into account the
complexity of the computation, it is difficult to suppose what the reason for
this disagreement could be.

\begin{figure}
\centerline{ \mbox{ \epsfxsize=0.5\textwidth \epsffile{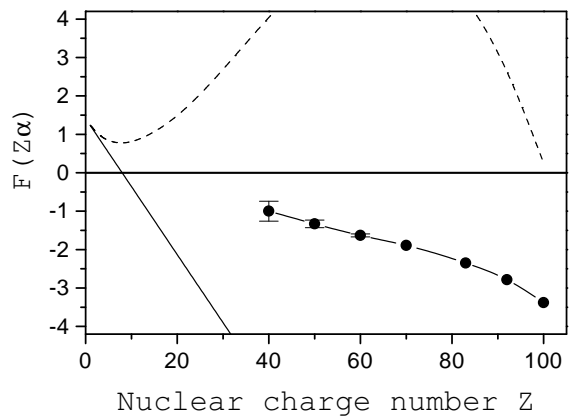} }} \caption{
The results of our numerical evaluation to all orders to $Z\alpha$ (dots)
together with the first two terms of the $Z\alpha$ expansion (solid line) and
all known terms of the $Z\alpha$ expansion (dashed line). \label{comparison}}
\end{figure}

As in the case of the one-loop self-energy, the evaluation becomes
problematic very fast as $Z$ decreases. It is due in part to the fact that
some individual contributions exhibit a nearly $Z$-independent behaviour,
while the total correction scales as $(Z\alpha)^4$. Numerical problems
restricted our calculation to the region $Z \ge 40$. In
Fig.~\ref{comparison}, we compare our non-perturbative results with the known
terms of the $Z\alpha$ expansion. Although we can not as yet say anything
conclusive about the higher-order terms, the figure suggests that the results
obtained by two different methods could be compatible.

Summing up, we have evaluated all contributions to the two-loop self-energy
correction for H-like ions with $Z\ge 40$. As this correction has been the
major source of the uncertainty of theoretical values for the ground-state
Lamb shift in these systems, our evaluation improves their accuracy by an
order of magnitude \cite{Yerokhin01sese}. While the experimental precision
for H-like uranium is not presently sufficient to probe the new contribution,
this should become possible when the experiment currently planned at GSI
\cite{Stoehlker00} is completed. The question of utmost importance is to
extend the present evaluation to low-$Z$ ions, where higher-order terms could
enter at the level of experimental interest even at $Z=$1 \cite{Pachucki01},
as well as to excited states. For the $2p_{1/2}$-$2s$ transition in Li-like
high-$Z$ ions the two-loop self-energy presently defines the uncertainty of
the theoretical prediction \cite{Yerokhin01sese} and can be probed by
comparing with experimental data available.

This study was supported in part by RFBR (Grant No. 01-02-17248), by Ministry
of Education (Grant No. E02-3.1-49), and by the program "Russian
Universities" (Grant No. UR.01.01.072). V.Y. acknowledges the support from
the Minist\`ere de l'Education Nationale et de la Recherche, the foundation
"Dynasty", and International Center for Fundamental Physics. The computation
was performed on the CINES and IDRIS national computer centers.

%

\newpage

\begin{table}
\caption{Individual contributions to the two-loop self-energy correction
expressed in terms of $F(Z\alpha)$.}
\begin{tabular}{lr@{}lr@{}lr@{}lr@{}lr@{}l}
$Z$  &  \multicolumn{2}{c}{LAL}
              &  \multicolumn{2}{c}{$F$ term}
                             &  \multicolumn{2}{c}{$P$ term}
                                      &  \multicolumn{2}{c}{$M$ term}
                                               &  \multicolumn{2}{c}{Total}
\\ \hline
 40 &  $-$0&.871  & 19&.50     &      $-$30&.13(15)  &    10&.50(18)   &    $-$1&.00(26)   \\
 50 &  $-$0&.973  & 10&.03     &      $-$14&.42(7)   &     4&.04(7)    &    $-$1&.33(10)   \\
    &      &      & 10&.02$^a$ &           &         &      &          &        &          \\
 60 &  $-$1&.082  &  5&.72     &       $-$7&.48(4)   &     1&.21(2)    &    $-$1&.63(4)    \\
 70 &  $-$1&.216  &  3&.497    &       $-$4&.03(3)   &  $-$0&.14(1)    &    $-$1&.89(3)    \\
    &  $-$1&.216$^b$& &        &       $ $ &         &      &          &        &          \\
 83 &  $-$1&.466  &  1&.938    &       $-$1&.831(13) &  $-$0&.990(5)   &    $-$2&.349(14)  \\
    &      &      &  1&.937$^a$&           &         &  $-$1&.66(1)$^a$&        &          \\
 92 &  $-$1&.734  &  1&.276    &       $-$1&.030(9)  &  $-$1&.295(3)   &    $-$2&.784(10)  \\
    &  $-$1&.733$^b$&1&.274$^a$&           &         &  $-$1&.855(7)$^a$&       &          \\
100 &  $-$2&.099  &  0&.825    &       $-$0&.635(6)  &  $-$1&.473(3)   &    $-$3&.382(7)   \\
    &      &      &  0&.825    &           &         &      &          &        &
\\ \hline
\end{tabular}

$^a\ $  Ref. \cite{Mallampalli98a}; $^b\ $  Ref. \cite{Mallampalli98}
\end{table}

\end{document}